\documentclass[apj]{emulateapj}
\usepackage{graphicx}
\usepackage{epsfig}
\usepackage{float}
\usepackage{lscape}

\newcommand{\hst}{{\it HST\/}}
\newcommand{\nclus}{NGC 6397}

\slugcomment{as of \today}
\shorttitle{47 Tuc and NGC 6397}
\shorttitle{Richer et al.}

\begin{document}

\title{Comparing the White Dwarf Cooling Sequences in 47 Tuc and NGC 6397}
\author{Harvey~B.~Richer\altaffilmark{1},  Ryan~Goldsbury\altaffilmark{1}, Jeremy~Heyl\altaffilmark{1},  Jarrod~Hurley\altaffilmark{2},
  Aaron~Dotter\altaffilmark{3},  
  Jason~S.~Kalirai\altaffilmark{4}, Kristin~A.~Woodley\altaffilmark{5},
    Gregory~G.~Fahlman\altaffilmark{6},
    R.~Michael~Rich\altaffilmark{7}, and Michael~M.~Shara\altaffilmark{8}} 

\altaffiltext{1}{Department of Physics \& Astronomy, University of
  British Columbia, Vancouver, BC, Canada V6T 1Z1; richer@astro.ubc.ca,  rgoldsb@phas.ubc.ca, heyl@phas.ubc.ca } 
   \altaffiltext{2}{Centre for Astrophysics \& Supercomputing, Swinburne
  University of Technology, Hawthorn, VIC 3122, Australia ;
  jhurley@swin.edu.au} 
  \altaffiltext{3}{Research School of Astronomy \& Astrophysics, Australian National University, Canberra, Australia;
  dotter@mso.anu.edu.au} 
     \altaffiltext{4}{Space Telescope Science Institute, 3700 San Martin
  Drive, Baltimore, MD, 21218; and Center for Astrophysical Sciences,
  Johns Hopkins University, Baltimore, MD, 21218;
  jkalirai@stsci.edu} 
\altaffiltext{5}{Department of Astronomy and Astrophysics, University of
  California at Santa Cruz, 1156 High Street, Santa Cruz, CA, 95064;
  kwoodley@ucolick.org}   
\altaffiltext{6}{National Research Council, Herzberg Institute of
  Astrophysics, Victoria, BC, Canada V9E 2E7;
  greg.fahlman@nrc-cnrc.gc.ca} 
    \altaffiltext{7}{Division of Astronomy and Astrophysics, University of
  California at Los Angeles, Los Angeles, CA, 90095; rmr@astro.ucla.edu} 
\altaffiltext{8}{Department of Astrophysics, American Museum of Natural
  History, Central Park West and 79th Street, New York , NY 10024;
  mshara@amnh.org} 
\begin{abstract}

Using deep Hubble Space Telescope imaging, color-magnitude diagrams are constructed for 
the globular clusters 47 Tuc and NGC 6397.  As expected, because of its lower metal abundance, the main sequence of NGC 6397 lies well to the blue
of that of 47 Tuc. A comparison of the white dwarf cooling sequences of the two clusters, however, demonstrates that these sequences are indistinguishable
over most of their loci - a consequence of the settling out of heavy elements in the dense white dwarf atmosphere and the near equality of their masses. Lower quality data on M4 continues this trend to a third cluster whose metallicity is intermediate between these two.
While the path of the white dwarfs in the color-magnitude diagram is nearly identical in 47 Tuc and NGC 6397, the numbers of white dwarfs along the path are not.  This results from the relatively rapid relaxation in NGC 6397
compared to 47 Tuc and provides a cautionary note that simply counting objects in star clusters in random locations as a method of testing stellar evolutionary theory is 
likely dangerous unless dynamical considerations are included.

\end{abstract}

\keywords{globular clusters:\ individual (NGC 6397, 47 Tuc) --- Stars:\ Population II --- Hertzsprung-Russell and C-M
  diagrams --- White Dwarfs}

%
\section{Introduction}
\label{sec:intro}

The majority of white dwarfs (WDs) have no metals in their atmospheres; their spectra indicate the presence only of  H and/or  He. This is in line with theoretical expectations (Vauclair et al. 1979) where
heavy elements in cooler WDs are expected to sink rapidly under the strong gravity of the WD. For most WDs, the star's location in the color-magnitude diagram (CMD) is not dependent on the metal abundance
of its progenitor.  

In addition to the locus of the WD sequence in clusters of different metallicity, the number distribution down the cooling sequence is also of interest. This could provide
information on a cluster's initial mass function or on WD physics such as the slow-down in
cooling when the WD core undergoes crystallization (van Horn 1968). Also, it could be a window on the dynamical processes internal to the cluster. Young WDs have recently changed their masses by about 50\%. Hence they should be in the process of altering their radial distribution and kinematics via relaxation processes (e.g. Spitzer \& Hart 1971, McLaughlin et al. 2006) and it may be possible to observe this.

In this contribution we examine the location and number distribution in detail in the CMDs of WDs in two globular star clusters of very different metallicities; NGC 6397 and 47 Tuc ([Fe/H] = $-$2.0 and $-$0.75, respectively (Carretta et al. 2009)). Using somewhat lower quality data, we also
include M4 in a comparison of the WD loci. The number of WDs down the cooling sequences
of 47 Tuc and NGC 6397 are explored, found to be quite different and then compared with the results of  $N-$ body calculations which suggest that internal dynamics has had an important role in controlling these distributions.

\section{Data and CMDs}

NGC 6397 is one of the closest globular clusters (2.6 kpc, see Richer et al.  2008 for a summary), and 47 Tuc is
a little less than twice as far from us (4.6 kpc, Woodley et al. 2012). For both clusters we made 
use of the superb imaging capabilities of the
Advanced Camera for Surveys (ACS) on the Hubble Space Telescope (HST) to explore their complete stellar populations.
For NGC 6397 we already had deep imaging data from previous work (Richer et
al.\ 2006, Hansen et al. 2007, Anderson et al. 2008), based on a large
program in \hst\ Cycle 13 (GO-10424, 126 orbits). This was supplemented by
a new epoch of observations in 2010 (GO-11633, 9 orbits)  in a single filter (F814W) in the same field
that was used to proper motion clean the earlier data.

\begin{figure*}
\epsfig{file=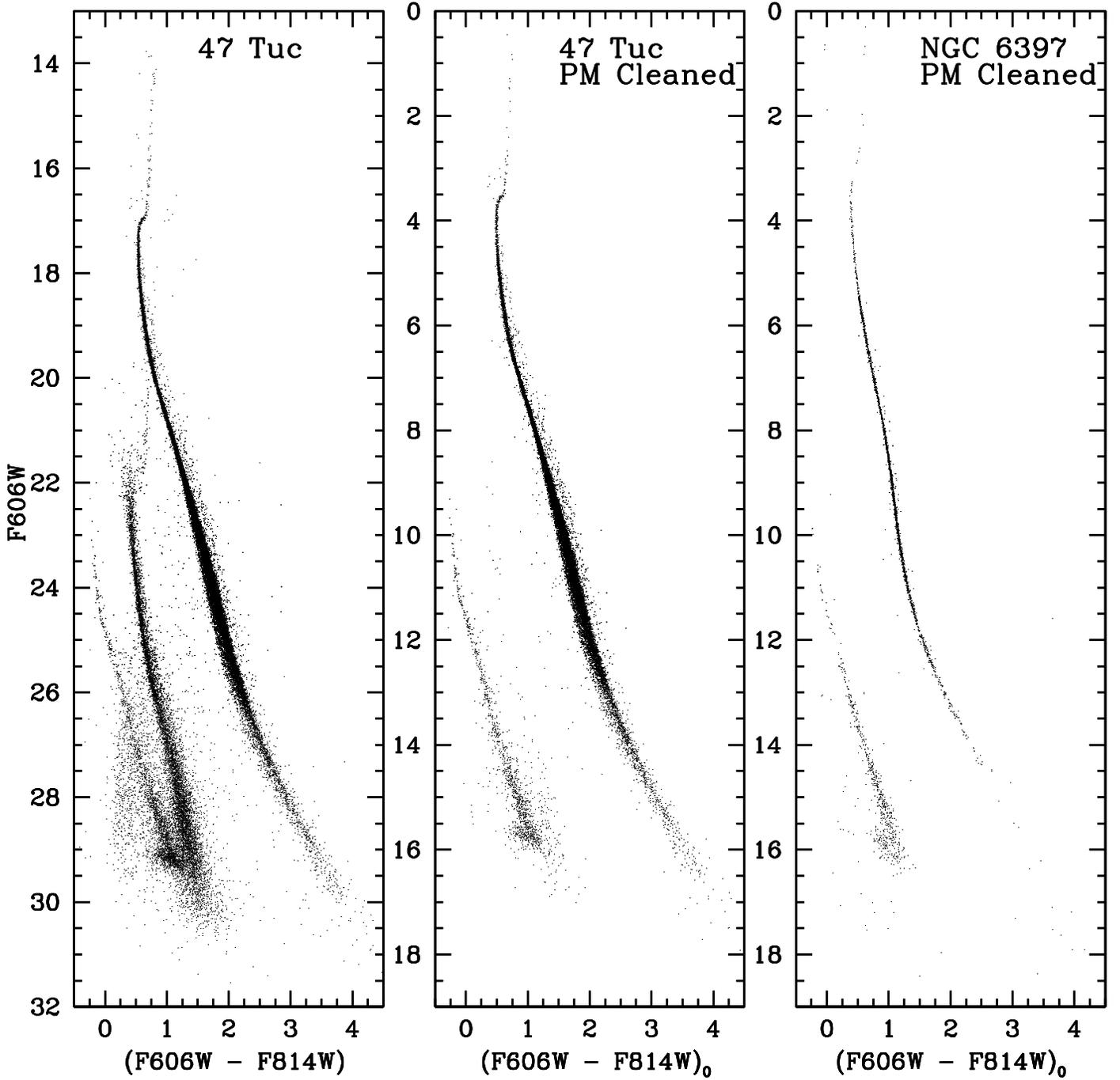,width=1.05\linewidth,clip=}
\caption{CMDs of 47 Tuc and NGC 6397. Left: All objects measured in our
  47 Tuc field (at 1.9 half light radii) that are more than 2.5$\sigma$ above the sky and that
  pass a modest error cut are included. The main sequence of the SMC, which is located in the background of 47 Tuc, lies between the 47 Tuc main sequence and its white dwarf cooling sequence. 
  In addition (middle panel) these data were proper motion cleaned so as to include just the cluster stars.  The axes in this plot are
  absolute magnitude and intrinsic color.  Right: Proper motion 
  cleaned CMD of NGC 6397 (at 1.7 half light radii). F606W apparent distance moduli of $13.42$ and
  $12.61$ are used for 47 Tuc and NGC 6397 respectively with their
  associated reddenings of $0.045$ and $0.223$. }
\label{fig:cmd}
\end{figure*}

For 47 Tuc our team was awarded
121 orbits in Cycle 17 to image the cluster (GO-11677).  The main science goal was to obtain
photometry with the ACS F606W and F814W filters that would go deep
enough to study the entirety of the white dwarf cooling sequence (Hansen et al. 2013). A detailed discussion of the observations and data
reduction can be found in Kalirai et al. (2012).

From these observations, we derived 
CMDs for the two clusters.
In the present paper we use for 47 Tuc a CMD constructed from stacks of
the ACS images in the F606W and F814W filters.  Our CMD is largely free of distracting artifacts; with
such clean images, the CMD penetrates to $\rm F606W>30$, even in a
field as crowded as this one. The left panel of Figure 1 displays the 47
Tuc CMD, showing all objects that were at least 2.5$\sigma$ above the
sky in both filters, and whose photometry fell within an error cut that
grew from 0.037 mag at $\rm F606W=20$ to 0.15 mag at $\rm F606W=30$.
In addition, these data were proper motion cleaned using archival images taken between 2002 and 2012 (see Richer et al. 2013 for details).

The far right panel of Figure 1 displays the PM-cleaned CMD for NGC 6397. Because we will want to compare the clusters directly,
the CMDs  have been plotted as
absolute magnitude and intrinsic color in the two right-most panels. To accomplish this, we begin with literature values for true distance moduli,
reddening and the ratio of total-to-selective absorption $\rm R_V$ (taken as 3.1 (Cardelli et al. 1989) for both clusters).  The absolute distance moduli
 for 47 Tuc and NGC 6397, respectively, are $13.30 \pm
0.08$ and $12.07 \pm 0.06$ with reddening values in ($\rm B -
\rm V$) of $0.04 \pm 0.01$ and $0.18 \pm 0.02$. This 47
Tuc distance was obtained from a mean of those discussed in Woodley et al. (2012) excluding their WD derived distance
(in order to avoid bias later when we compare the cooling sequences of the cluster WDs) and the NGC 6397 values are from Richer et al. (2008)
again excluding WD derived values.

We now need to convert these to $\rm F606W$ and $\rm F814W$ and apply the extinction correctly in our filters. 
Because we are interested mainly in the WDs in this contribution, we determine the extinction corrections for stars with average WD temperatures.
For this we use the spectral model 
grid from Tremblay et al. (2011) to determine a model 
spectrum for a 10,000K WD which falls near the centre of
our WD cooling sequences in these clusters.  We then use the reddening curve from 
Fitzpatrick (1999) to calculate the magnitudes of this
object in our filters with and without interstellar
reddening applied.  We use the differences of these 
magnitudes to calculate the appropriate extinction in
each filter and correct our CMDs to an unreddened
frame.  We then use true distance moduli to adjust each 
CMD so that it lies in an extinction and distance 
corrected reference frame. 
    
 \section{Main Sequence and White Dwarf Loci}
\label{sub:loci}

It is not immediately obvious from Figure 1, but the MSs of the two clusters have rather different colors.  This
becomes evident when we overplot the two CMDs, as in Figure 2, where the CMD of \nclus\ was plotted in red, and
that of 47 Tuc was then overplotted in black.  At the brightness level of the hot WDs, the MSs
 differ in color index by about half a magnitude.  This is of course due
to the large difference between the metallicities of the two clusters; it has been recognized for
many decades that MS color is controlled by the amount of
line blanketing that metals produce in a stellar atmosphere (e.g.  Sandage \& Walker 1966).

 \begin{figure}
\epsfig{file=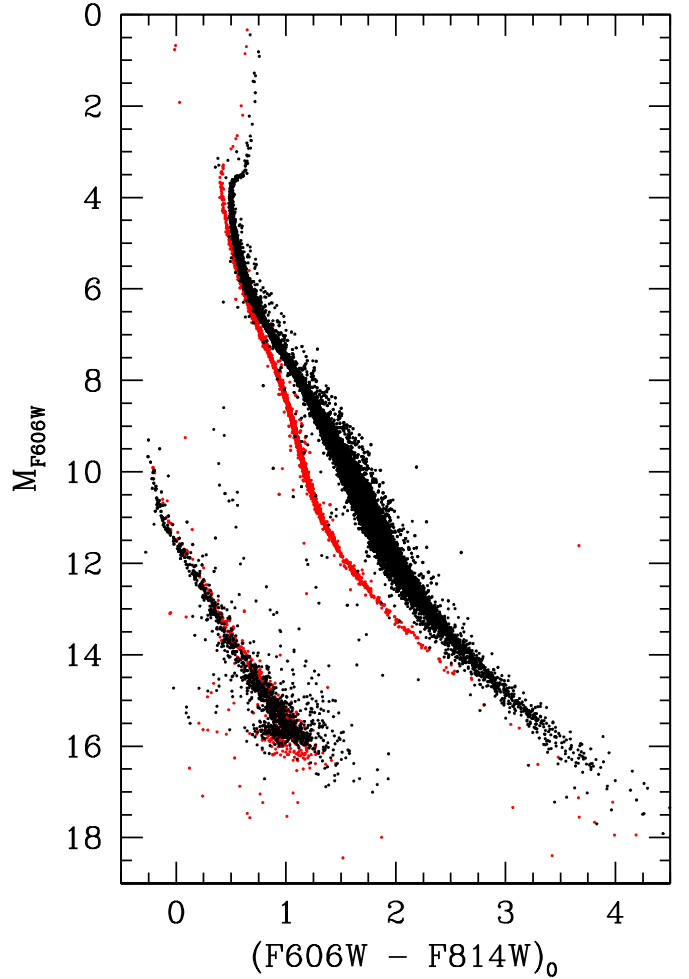,width=\linewidth} 
\caption{The entire CMDs of NGC 6397 (red) over-plotted with that of 47 Tuc (black) in absolute magnitude and intrinsic color. The MSs of the clusters do not align while the WD cooling sequences do.}
\label{fig:overlay}
\end{figure}

The difficulty in using MS turnoffs for age determination becomes obvious in this diagram: the large difference between turnoff locations (which
eventually determines the age) could be due either to a metallicity and/or an age difference. By contrast, the expectation is that the WD cooling sequences of the clusters should align, independent of the metallicity of the progenitor. The WDs should have no
metals in their atmospheres (Fontaine \& Michaud 1979). These sink rapidly under the strong gravity of the WD unless the star is very hot and can radiatively levitate atoms (Holberg et al. 1994, Chayer et al. 1995), or
unless the atmosphere is continually being polluted from circumstellar material (e.g. Farihi et al. 2009).  Hence their spectra and location in the CMD do not explicitly depend on the cluster metal abundance - except for a possible
WD mass dependence on metallicity through the initial-final mass relation (Kalirai et al. 2008, 2009).  This point was discussed in detail in Woodley et al. (2012), but there is no evidence
for a range in the masses of, for example,  globular cluster WDs with metallicity at the top of the cooling sequence (Moehler et al. 2004, Kalirai et al. 2009) although the 
data here are not yet conclusive. 

Over the full range of the
WD cooling sequences the overlap between the clusters is remarkable (see Hansen et al. (2013) Figure S5 for an earlier version of this diagram) and we demonstrate below
that this result is very robust to modest changes in distance modulus and reddening.

To explore the location of the WD loci in more detail, in Figure 3 we include M4, the only other globular cluster with data deep enough to see a major component of the cooling sequence. Its metallicity ([Fe/H] $=$ $-$1.20 (Drake et al. 1994)) lies between that of 47 Tuc and NGC 6397. The data for this cluster were downloaded 
from the HST archives. The filters  used here were F606W and F775W so we were required
to transform the color to (F606W $-$ F814W). Using a cooling model from Fontaine et al. (2001) and spectral models from Tremblay et al. (2011) we
found a relation between temperature and reddened (F606W $-$ F775W) color (using E(B$-$V) $=$ 0.35 (Richer et al. (1997) and R$_{V}$ $=$ 3.76
(Kaluzny et al. (2013)) for WDs of 0.53M$_{\odot}$.
We then calculated (F606W $-$ F814W) for the same temperatures.  This allows us to determine an expected difference between these two colors as a function of reddened (F606W $-$ F775W).  This color dependent transform is then applied to our original photometry to generate a transformed (F606W $-$ F814W) CMD.  
These M4 data are of poorer quality and the
scatter at the faint end is much larger. Using the true distance modulus from Kaluzny et al. (2013) of 11.34,
we plot the M4 stars in Figure 3 with smaller points, so all cluster sequences are readily visible.
We collect the major properties of the clusters we have been discussing in Table 1.

 \begin{deluxetable}{llc}[H!]
\tablewidth{0pc}
\tablecaption{Cluster Properties  \label{table1A}}
 \tablehead{
\colhead {Parameter} & \colhead{Value} & \colhead{Reference} 
 }
\startdata
\\
&47 Tuc & \\
true distance modulus & 13.30 $\pm$ 0.08 & 1\\
$\rm E(B-V)$ & 0.04 $\pm$ 0.01 & 1 \\
$\rm E(F606W - F814W)$ & 0.045 & 2 \\
$\rm R_V$ & 3.1 & 3\\
$\rm (m - M)_{F606W}$& 13.42 & 2 \\
$\rm [Fe/H]$ & $-0.75$ & 4\\
\\
&NGC 6397&\\
true distance modulus & 12.07 $\pm$ 0.06 & 5\\
$\rm E(B-V)$ & 0.18 $\pm$ 0.02 & 5 \\
$\rm E(F606W - F814W)$ & 0.223 & 2 \\
$\rm R_V$ & 3.1 & 3\\
$\rm (m - M)_{F606W}$& 12.61 & 2 \\
$\rm [Fe/H]$ & $-2.00$ & 4 \\
\\
&M4&\\
true distance modulus & 11.34 $\pm$ 0.03 & 6\\
$\rm E(B-V)$ & 0.35 $\pm$ 0.02 & 7 \\
$\rm E(F606W - F814W)$ & 0.45 & 2 \\
$\rm R_V$ & 3.76 & 6\\
$\rm (m - M)_{F606W}$& 12.57 & 2 \\
$\rm [Fe/H]$ & $-1.20$ & 8 \\

  \enddata
\tablerefs{(1) Woodley et al. 2012; (2) This paper; 
(3) Cardelli et al. 1989; (4) Carretta et al. 2009; (5) Richer et al. 2008; 
(6) Kaluzny et al. 2013; (7) Richer et al. 1997; (8) Drake et al. 1994}
\end{deluxetable}

 \begin{figure}
\plotone{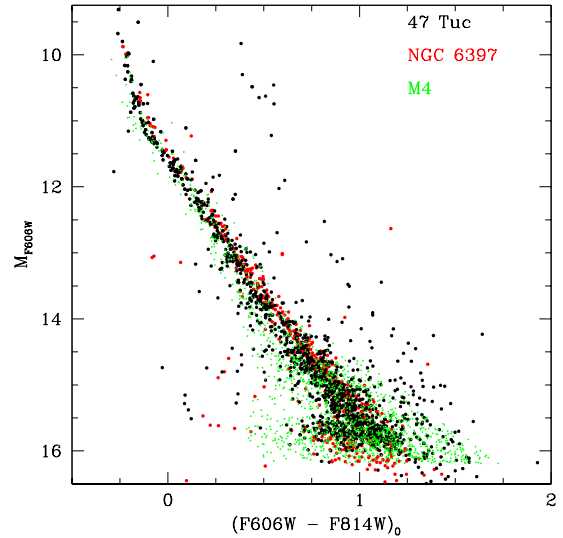}
\caption{ An overlay  of the WD cooling sequences in 47 Tuc, NGC 6397 and M4.}
\label{fig:overlay}
\end{figure}

We compared a bright portion of the WD cooling sequences between $\rm M_{F606W} =11.5-15.0$ for all three clusters in order to assess whether their slopes and intercepts were consistent.
This region avoids the brightest WDs whose insulating  hydrogen layers are not fully degenerate as well as the fainter WDs where the photometry becomes less certain.  We used a maximum likelihood method to fit a line to each WD sequence. Our fitting method also varies distance and reddening.  At the end we marginalize these parameters out with priors that reflect a 
20\% uncertainty in reddening/extinction and a 0.1 magnitude uncertainty in true distance modulus.  The result is a likelihood surface for only the two model parameters (slope and intercept) for each cluster (Figure 4). These contours reflect the inherent scatter in the photometry, the photometric errors, and uncertainties in distance and extinction. The 1$\sigma$ contours overlap for all three clusters demonstrating the near equality 
of the loci of the WD cooling sequences for clusters covering almost the entire range of metallicity in the globular cluster system.

 \begin{figure}
\plotone{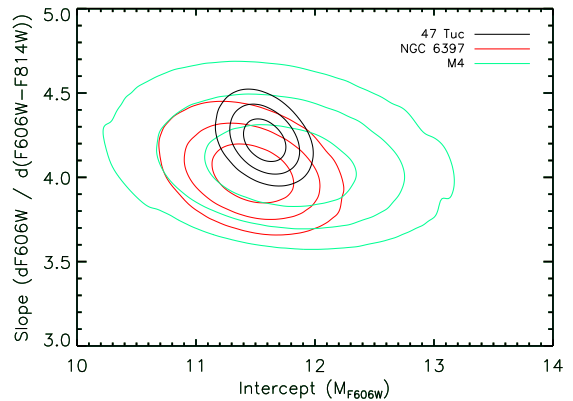}
\caption{Significance contours for slope and intercept of the WD cooling sequences in the three clusters. The contours are 1, 2 and 3$\sigma$.}
\label{fig:contour}
\end{figure}

Turning now to the full WD cooling sequences, we note that these are well populated in all three clusters and show the theoretically predicted  turn to bluer colors at faint magnitudes (Bergeron et al. 1995, Hansen 1999, Saumon \& Jacobson 1999). This is due to collision-induced absorption which, in the cool dense atmosphere of such a WD, allows H$\rm _2$ to form with its strong IR opacity. This has the result of forcing the radiation out in the bluer spectral regions so the star gets bluer even though its temperature is getting cooler. 

The location of the truncation of the WD cooling sequence is a potentially sensitive age indicator as it is the limit to which the bulk of the WDs could have cooled over the lifetime of the cluster. Older clusters are thus expected to exhibit fainter truncation magnitudes, as is seen for NGC 6397
compared to 47 Tuc.  No strong conclusions can be made for M4 considering the large scatter in the data and the lack of incompleteness corrections in this plot.
A detailed comparison employing
incompleteness corrections for these very faint stars and a full modelling of the WD cooling sequences for 47 Tuc and NGC 6397 has been carried-out (Hansen et al. 2013). NGC 6397 is found to be 2 Gyr older than 47 Tuc.

\section{Comparing Numbers of WDs}
\subsection{Using Cooling Model Ages}

The number of WDs as a function of magnitude (the luminosity function) contains  information about the cluster age, its  initial-mass function, WD cooling physics and cluster dynamics. This motivates a detailed comparison of the number distribution with magnitude of WDs in 47 Tuc and NGC 6397.

To explore this quantitatively, we analyze the numbers of WDs of various ages down the cooling sequence. 
To accomplish this we compare the completeness corrected counts of WDs in various magnitude ranges (that is, cooling ages)
to MS stars with masses similar to the WDs (0.5-0.6 $\rm M_{\odot}$) in the two clusters. This mass
was chosen to match that of the WDs so that any differences could be most easily interpreted.
 The ages of the WDs were obtained from the cooling models of Fontaine et al. (2001). The data are contained in Table 2. Statistical errors based on the numbers of stars counted are
included in the brackets beside each entry and the age of the WD sample is indicated.

 \begin{deluxetable}{lccc}
\tablecolumns{4}
\tablewidth{0pc}
\tablecaption{Ratio WDs of Various Ages to 0.5 $-$ 0.6 M$_{\odot }$ MS Stars \label{table1}}
\tablehead{
\colhead{Cluster} & \colhead{WD $<$0.5 Gyr} & \colhead{0.5 -- 2 Gyr} &
        \colhead{ 2 -- 5 Gyr}  
 }
\startdata
47 Tuc & 0.047(0.003) & 0.071(0.004) & 0.114(0.004) \nl
N6397 & 0.063(0.013) & 0.278(0.026) & 0.714(0.039) \nl
\vspace{0.1mm} \\
\hline
\vspace{0.1mm} \\
Model \nl
 $t/t_{rh} = 2.5$ & 0.038(0.008) & 0.086(0.011) & 0.180(0.021)  \nl
$t/t_{rh} = 35$ & 0.062(0.012) & 0.248(0.029) & 0.564(0.022)  \nl
 \enddata
 \end{deluxetable}

For WDs younger than about 0.5 Gyr, the WD/MS ratio is similar in the two clusters. For WD ages greater than this the difference between the ratios grows so that for
5 Gyr WDs it is more than six times larger in \nclus\ than in 47 Tuc.

There are a number of potential reasons for these differences. Age
is one possibility. Hansen et al. (2013) have shown that 47 Tuc is younger by about 2 Gyr. The number of WDs will increase with age in a cluster because of stellar evolution,
roughly doubling for a factor of two in age. However, the small age difference between 47 Tuc and NGC 6397 is insufficient 
for this to be the full answer.
Others causes for the variations in these ratios may be binary fraction and
the cluster orbit, both possibilities could have an effect on the WD/MS ratio by hiding WDs in the first instance and stellar evaporation in the second case.
We also know that there is a metallicity difference and that could
play a role.
But the biggest difference is likely to be dynamical age.
NGC 6397 is post-core-collapse with a half-mass relaxation time ($t_{rh}$) of 0.4 Gyr (Harris 2010)
while 47 Tuc is listed in the same reference
as non-core-collapse with $t_{rh} \sim$ 3.5 Gyr. So the age/$t_{rh}$ ($t/t_{rh}$) values
are very different for the two clusters; $\sim 30$ for NGC 6397 and $\sim 3$ for 47 Tuc.

   \begin{deluxetable*}{lccc}
\tablecolumns{6}
\tablewidth{0pc}
\tablecaption{WD Numbers at Various Ages \label{table2}}
\tablehead{
\colhead{Cluster} & \colhead{M$\rm _{F606W}$ $<$11.63} & \colhead{11.63 $<$ M$\rm _{F606W}$ $<$13.10} &
     \colhead{13.10 $<$ M$\rm _{F606W}$ $<$14.15} 
 }
\startdata
WD Cooling Age (Gyr) & $<$0.7 & 0.7$-$2.1  & 2.1$-$3.5 \nl
 Number 47 Tuc WDs  & 74 & 148 & 148  \nl
 Number NGC 6397 WDs & 20 & 49 & 81 \nl
\enddata
 \end{deluxetable*}

For further insight here, we examined a number  of $N-$body star cluster models (see Hurley et al. (2008), Hurley \& Shara (2012) for details of the modelling approach) that included stellar evolution, a Galactic tidal field, binaries and their  formation and destruction.
We use these models mainly to investigate trends as opposed to using them to predict actual numbers of stars. 

One model in particular is, by design, a good match for NGC 6397. The orbit, binary fraction and initial compactness were chosen to match the cluster. This model initially consisted of 200,000 stars with 2.5\% binaries and, at an age of 11.7 Gyr, has $t/t_{rh} = 35$. By this time it has lost more than 
80\% of its stars. By contrast, NGC 6397 currently has about 200,000 stars (Heyl et al.  2012). The ratios of WDs of various ages to MS stars
with masses  between 0.5 and 0.6 M$_{\odot }$ were extracted from this model between 1.2 and 2.2 half-light radii, similar to our observed field in NGC 6397. These ratios, with associated errors based on the numbers of model stars counted, are contained in the final row of Table 2. The increase in the WD/MS ratio here is dramatic, nicely similar to that seen in NGC 6397
and quite different from the behavior observed in 47 Tuc.

Regarding 47 Tuc, we do not have a model that comes even close to being a real representation of this cluster. It currently has about 1 million stars and likely began life with double this number (Giersz \& Heggie 2011). Such
a population is still well beyond the capabilities of $N-$body calculations. In the models we have 
calculated thus far, clusters with small $t/t_{rh}$ are generally young. In the same $N$ $=$ 200,000 model as above, the cluster has $t/t_{rh} = 3$ when it is only 4 Gyr old. Because of stellar evolution, this would 
have reduced the total fraction of WDs in the cluster by at least a factor of two so any comparison with the real cluster would be difficult. In addition, these WDs are more massive than those currently being produced in the real cluster. Our best match to 47 Tuc was a model that began with $N$ $=$ 100,000 stars, with quite a different relaxation history from the 200,000 model in that at about 13 Gyr it had $t/t_{rh} = 2.5$.
This model began with 5\% primordial binaries. By 13 Gyr, $N$ had been reduced to 79,500 stars, a far cry from the million stars still present in 47 Tuc today. For this model cluster the stellar statistics were 
extracted between 1.4 and 2.4 half-light radii, similar to the location of our real field in 47 Tuc. These are contained in the second-to-last row of Table 2. Here the WD/MS variation with age is much more gradual, more in line with
the numbers seen in 47 Tuc.

 Since the real clusters don't differ in age by more than 20\% (Hansen et al. 2013), and since the loss of stars via evaporation would decrease both the MS and WD numbers, it appears from
 a comparison with the $N-$body models that internal dynamical evolution is the dominant contributor to the WD/MS trends observed in these two clusters. The initial mass functions of the two model clusters were identical Kroupa functions (Kroupa et al. 1993) so the  variations seen in the models are not due to this quantity.
Most of the precursors to the WDs currently seen in our NGC 6397  field were likely
born elsewhere (presumably closer to the core) and they migrated into our field after they lost much of their mass on becoming WDs. Within this scenario, the much shorter half-mass relaxation time in NGC 6397 (0.4 Gyr) versus 3.5 Gyr for 47 Tuc, is 
consistent with the ratios in the two clusters being similar for WDs younger than 0.5 Gyr and growing more rapidly in NGC 6397 relative to 47 Tuc for older WDs.

 \subsection{Using Statistical Ages}
  
In the previous subsection we relied on models to set  the WD cooling timescale and $N-$body calculations to help interpret the results. Here the
discussion avoids using theoretical models, instead using a statistical approach to estimate cooling ages.

The rate of formation of WDs in our 47 Tuc ACS field is  1/0.0095 WD/Gyr (Goldsbury et al. 2012). This comes from counting the giant stars in our field, establishing their evolutionary time through theoretical models (giant models from Dotter et al. 2008, not WD models) and assuming that all the giants eventually turn into WDs. The relaxation time in our NGC 6397 field is 0.7 Gyr (Heyl et al. 2012). Note that this is larger
than the half-mass relaxation time as the field is farther from the cluster center.
We divide the hot end of the 47 Tuc WD cooling sequence into three sections by age; stars younger than 0.7 Gyr, stars between 0.7 and 2.1 Gyr, and 2.1 to 3.5 Gyr. The number of WDs less than
0.7 Gyr is 0.7/0.0095 $=$ 74. Because the relaxation time in our 47 Tuc field is at least 3.5 Gyr (Harris 2010), we do not expect significant numbers of
 stars to arrive or leave this field, on average, in a single relaxation time, so the WD cooling age across each bin should scale as the number of WDs in that bin.
 Hence if we proceed down the 47 Tuc WD cooling sequence until we have counted 74 stars, that will correspond to a cooling time of 0.7 Gyrs for the faintest object. Counting down to $3 \times 74$ stars would be at an absolute magnitude that corresponds to a cooling time of 2.1 Gyr, with $5\times 74$
 stars at a cooling age of 3.5 Gyr. The assumption is made here that the formation rate of WDs has been constant over the last few Gyr. Goldsbury et al. (2012) have shown that this assumption is valid in our 47 Tuc field.

 Since the WD cooling rate is not dependent on metallicity, and since the WD cooling sequences align so that there is little if any mass difference, the same absolute magnitude ranges must also correspond to WDs at the same cooling ages in NGC 6397. We now examine
 what the incompleteness-corrected WD numbers are in these increasingly faint magnitude bins for NGC 6397. The counts are contained in Table 3. In the first NGC 6397 bin (down to an absolute F606W magnitude of 11.63), we 
 count 20 WDs. In the next bin we expect 40 (the bin is twice as wide in age) and count 49. The Poisson probability of this occurring is 3\%. In the final bin we count 81 where, again, 40 are expected. 
 The probability of this occurrence by chance is 1.8 $\times 10^{-8}$.
 
For NGC 6397 the WD numbers grow as we progress to the fainter bins. Since the two clusters don't differ in age by a large amount and since evaporative losses would likely
be more important in NGC 6397 than in 47 Tuc, it again emerges that the NGC 6397 cooling sequence is being populated not just by stars evolving in the observed field, but large numbers must be coming into our field
from elsewhere.

\section{Conclusions}

The main result of this paper is a demonstration that WD cooling sequences for clusters of very different metallicities align almost perfectly in the CMD. This is in line with theoretical expectations
where heavy elements in the dense atmosphere of a WD are expected to sink, yielding atmospheres composed solely  of H and/or He. Virtually all knowledge of the original metal abundance of the WD-precursor has been lost. This is a strong endorsement of the use of WDs in cosmochronolgy where the dependence on metallicity found in other dating techniques (particularly MS turnoffs) are largely
eliminated.

Most of the WDs found in our NGC 6397 field, which is located near two half-mass radii in the cluster, likely formed nearer to the cluster center and have slowly been migrating outward. 
A red giant  with a mass near 0.8 ${\rm M_{\odot}}$ located close to the cluster center evolves into a 0.5 ${\rm M_{\odot}}$ WD. Two$-$body relaxation drives the star to a larger radius orbit. Because of the relatively
short relaxation time in NGC 6397, over a period of a few Gyr these WDs end up increasing the ratio of WDs to MS stars of the same mass at radii beyond the core. It appears that we are seeing this process in action in our data.

\medskip
Based on observations with the NASA/ESA Hubble Space Telescope, obtained at the Space Telescope Science Institute, which is operated by the Association of Universities for Research in Astronomy, Inc., under NASA contract NAS5-26555. These observations are associated with proposal GO-11677. Support for program GO-11677 was provided by NASA through a grant from the Space Telescope Science Institute which is operated by the Association of Universities for Research, Inc., under NASA contract NAS5-26555. H.B.R. and J.H. are supported by grants from The Natural Sciences and Engineering Research Council of Canada and by the University of British Columbia. J.S.K., R.M.R and M.M.S. were funded by NASA. A.D. received support from the Australian Research Council under grant FL110100012.

\newpage



\begin{thebibliography}{}
 \bibitem[Anderson et al.(2008)]{anderson08} Anderson, J., King, I.~R.,
  Richer, H.~B., Fahlman, G.~G., Hansen, B.~M.~S., Hurley, J., Kalirai,
  J.~S., Rich, R.~M., \& Stetson, P.~B. 2008, \aj, 135, 2114  
   \bibitem[Bergeron et al.(1995)]{bergeron95} Bergeron, P., Wesemael,
  F. \& Beauchamp, A. 1995, \pasp, 107, 1047 
  \bibitem[Cardelli1989]{card89} Cardelli, J.A., Clayton, G.G., Mathis, J.S. 1989, \apj, 345, 245
  \bibitem[Carretta et 
al.(2009)]{2009A&A...508..695C} Carretta, E., Bragaglia, A., Gratton, R., D'Orazi, V., \& Lucatello, S.\ 2009, \aap, 508, 695 
  \bibitem[Chayer et al.(1995)]{1995ApJS...99..189C} Chayer, P., Fontaine, 
G., \& Wesemael, F.\ 1995, \apjs, 99, 189 
  \bibitem[Dotter et al.(2008)]{2008ApJS..178...89D} Dotter, A., Chaboyer, 
B., Jevremovi{\'c}, D., et al. 2008, \apjs, 178, 89 
\bibitem[Drake et al.(1994)]{1994ApJ...430..610D} Drake, J.~J., Smith, 
V.~V., \& Suntzeff, N.~B.\ 1994, \apj, 430, 610 
\bibitem[Farihi et al.(2009)]{2009ApJ...694..805F} Farihi, J., Jura, M., 
\& Zuckerman, B.\ 2009, \apj, 694, 805 
    \bibitem[Fitzpatrick(1999)]{fitzpatrick99} Fitzpatrick, E.~L. 1999,
  \pasp, 111, 63 
  \bibitem[Fontaine 
\& Michaud(1979)]{1979ApJ...231..826F} Fontaine, G., \& Michaud, G.\ 1979, \apj, 231, 826 
  \bibitem[Fontaine et al.(2001)]{2001PASP..113..409F} Fontaine, G., 
Brassard, P., \& Bergeron, P. 2001, \pasp, 113, 409 
\bibitem[Giersz 
\& Heggie(2011)]{2011MNRAS.410.2698G} Giersz, M., \& Heggie, D.~C.\ 2011, \mnras, 410, 2698 
\bibitem[Goldsbury et al.(2012)]{2012ApJ...760...78G} Goldsbury, R., Heyl, 
J., Richer, H.~B., et al. 2012, \apj, 760, 78 
 \bibitem[Gratton et al.(2003)]{gra03} Gratton, R.~G., Bragaglia, A.,
  Carretta, E., et al. 2003, \aap, 408, 529
  \bibitem[Hansen(1999)]{hansen99} Hansen, B.~M.~S. 1999, \apj, 520, 680 
  \bibitem[Hansen et al.(2007)]{hansen07} Hansen, B.~M.~S., Anderson, J.,
  Brewer, J., Dotter, A., Fahlman, G.~G., Hurley, J., Kalirai, J., King,
  I., Reitzel, D., Richer, H.~B., Rich, R.~M., Shara, M.~M., \& Stetson,
  P.~B. 2007, \apj, 671, 380 
    \bibitem[Hansen et al.(2013)]{2013Natur.500...51H} Hansen, B.~M.~S., 
Kalirai, J.~S., Anderson, J., et al.\ 2013, \nat, 500, 51 
 \bibitem[Harris(2010)]{harris10} Harris, W.~E. 2010, arXiv1012.3224 
\bibitem[Heyl et al.(2012)]{2012ApJ...761...51H} Heyl, J.~S., Richer, H., 
Anderson, J., et al. 2012, \apj, 761, 51 
\bibitem[Holberg et al.(1994)]{1994ApJ...425L.105H} Holberg, J.~B., Hubeny, 
I., Barstow, M.~A., et al.\ 1994, \apjl, 425, L105 
\bibitem[Hurley et al.(2008)]{hurley08} Hurley, J.~R., Shara, 
M.~M., Richer, H.~B., et al. 2008, \aj, 135, 2129 
\bibitem[Hurley 
\& Shara(2012)]{2012MNRAS.425.2872H} Hurley, J.~R., \& Shara, M.~M.\ 2012, \mnras, 425, 2872 
\bibitem[Kalirai et al.(2008)]{kal08} Kalirai, J.~S., Hansen, 
B.~M.~S., Kelson, D.~D., et al. 2008, \apj, 676, 594 
\bibitem[Kalirai et al.(2009)]{kal09} Kalirai, J.~S., Saul 
Davis, D., Richer, H.~B., et al. 2009, \apj, 705, 408 
  \bibitem[Kalirai et al.(2012)]{kalirai12} Kalirai, J.~S., Richer, H.~B.,
  Anderson, J., et al. 2012, \aj, 143, 11
  \bibitem[Kaluzny et al.(2013)]{2013AJ....145...43K} Kaluzny, J., Thompson, 
I.~B., Rozyczka, M., et al.\ 2013, \aj, 145, 43 
\bibitem[Kroupa et al.(1993)]{1993MNRAS.262..545K} Kroupa, P., Tout, C.~A., 
\& Gilmore, G.\ 1993, \mnras, 262, 545 
\bibitem[McLaughlin et al.(2006)]{2006ApJS..166..249M} McLaughlin, D.~E., 
Anderson, J., Meylan, G., et al.\ 2006, \apjs, 166, 249 
   \bibitem[Moehler et al.(2004)]{moe04} Moehler, S., Koester, D., Zoccali,
  M., et al. 2004, \aap, 420, 515
  \bibitem[Richer et al.(1997)]{1997ApJ...484..741R} Richer, H.~B., Fahlman, 
G.~G., Ibata, R.~A., et al.\ 1997, \apj, 484, 741 
   \bibitem[Richer et al.(2006)]{richer06} Richer, H.~B., Anderson, J.,
  Brewer, J., Davis, S., Fahlman, G.~G., Hansen, B.~M.~S., Hurley, J.,
  Kalirai, J.~S., King, I.~R., Reitzel, D., Rich, R.~M., Shara, M.~M.,
  \& Stetson, P.~B. 2006, Science, 313, 936 
\bibitem[Richer et al.(2008)]{richer08} Richer, H.~B., Dotter, A.,
  Hurley, J., Anderson, J., King, I., Davis, S., Fahlman, G.~G., Hansen,
  B.~M.~S., Kalirai, J., Paust, N., Rich, R.~M., \& Shara, M.~M. 2008,
  \aj, 135, 2141 
   \bibitem[Richer et al.(2013)]{2013ApJ...771L..15R} Richer, H.~B., Heyl, J., 
Anderson, J., et al.\ 2013, \apjl, 771, L15 
  \bibitem[Sandage 
\& Walker(1966)]{1966ApJ...143..313S} Sandage, A., \& Walker, M.~F. 1966, \apj, 143, 313 
\bibitem[Saumon \& Jacobson(1999)]{saumon99} Saumon, D., \& Jacobson,
  S.~B. 1999, \apjl, 511, L107 
bibitem[Spitzer 
\& Hart(1971)]{1971ApJ...164..399S} Spitzer, L., Jr., \& Hart, M.~H.\ 1971, \apj, 164, 399 
   \bibitem[Tremblay et al.(2011)]{tremblay2011} Tremblay, P.-E.,
   Bergeron, P., \& Gianninas, A. 2011, \apj, 730, 128
   \bibitem[van Horn(1968)]{1968ApJ...151..227V} van Horn, H.~M.\ 1968, \apj, 
151, 227 
\bibitem[Vauclair et 
al.(1979)]{1979A&A....80...79V} Vauclair, G., Vauclair, S., \& Greenstein, J.~L.\ 1979, \aap, 80, 79 
\bibitem[Woodley et al.(2012)]{woodley12} Woodley, K.~A., et al. 2012,
  AJ, 143, 50
 
\end{thebibliography}
\end{document}